# Standardizing case study descriptions for multi-energy systems and networks modeling

*Mathieu Vallee[a], Eva Schischke[b], Edmund Widl[c], Gabriela Zabik[d], Mohamed-Tahar Mabouk[e], Kai Derzsi[f], Dirk Müller[f], Sergio Rech[g], Costanza Saletti[h], Martina Capone[i], Sacha Hodencq[j], Ildar Daminov[e], Gabriele Leoncini[a], Marc Clausse[k]*

[a] *Université Grenoble Alpes, CEA LITEN, DTCH, Le Bourget-du-Lac, France, mathieu.vallee@cea.fr*
[b] *Fraunhofer Institute for Environmental, Safety and Energy Technology UMSICHT, Oberhausen, Germany, eva.schischke@umsicht.fraunhofer.de*
[c] *Austrian Institute of Technology, Vienna, Austria Institution, City, Country, Edmund.Widl@ait.ac.at*
[d] *TU Wien, Institute of Energy Systems and Thermodynamics, Vienna, Austria, gabriela.zabik@tuwien.ac.at*
[e] *IMT Atlantique, Department of Energy Systems and Environment, GEPEA UMR CNRS, 6144, F-44307, Nantes, France, mohamed-tahar.mabrouk@imt-atlantique.fr*
[f] *E.ON Energy Research Center, Institute for Energy Efficient Buildings and Indoor Climate, RWTH Aachen University, Germany, kai.derzsi@eonerc.rwth-aachen.de*
[g] *University of Padova, Department of Industrial Engineering, Padova, Italy, sergio.rech@unipd.it*
[h] *University of Parma, Department of Engineering for Industrial Systems and Technologies, Parma, Italy, costanza.saletti@unipr.it*
[i] *Politecnico di Torino, Energy Department, Torino, Italy, martina.capone@polito.it*
[j] *Université Grenoble Alpes, G2ELab, Grenoble, France, sacha.hodencq@univ-grenoble-alpes.fr*
[k] *INSA Lyon, CNRS, CETHIL, UMR 5008, Villeurbanne, 69100, France, marc.clausse@insa-lyon.fr*

## Abstract:

Research on Multi-Energy Systems (MES) often relies on case studies with divergent hypotheses and terminologies, limiting comparability and slowing progress. Discussions at the ECOS 2025 conference highlighted the need for standardized reference case studies to facilitate reuse and comparison. While frameworks like the IEC 62559 standard and the Open Energy Platform (OEP) exist, their adoption for MES remains fragmented. This heterogeneity hinders collaboration and replicability, motivating efforts towards a unified description framework tailored to MES.

This paper aims to address this gap by evaluating existing approaches in order to promote a standardized description framework for MES case studies. The goal is to enhance comparability, streamline research, and make a first step towards defining reference case studies and benchmarks in the domain.

The study adopts a collaborative approach: after analysing existing description frameworks and selecting the most suitable one, the co-authors describe their own case studies, followed by cross-reviews to assess completeness, clarity, and openness of data/models. The description framework is adapted to emphasize MES-specific elements, such as system configuration and use case details. A checklist is developed to guide reviews.

Preliminary results include a set of standardized case study descriptions and insights from cross-reviews on framework strengths/limitations. The diversity of case studies underscores the framework's flexibility, while feedback reveals opportunities for improvement and broader adoption.

This work provides a foundation for standardized MES case study descriptions, fostering collaboration, comparability, and replicability. By reducing ambiguity and ensuring the availability of relevant information in a consistent format, it accelerates research and benchmarking in the field.

# 1. Introduction

Replication and reproducibility are some of the pillars of scientific research. According to [1], reproducible results are obtained when “the main results of the paper have been obtained in a subsequent study by a person or team other than the authors, using, in part, artifacts provided by the author”. Replication is obtained when ”the main results of the paper have been independently obtained in a subsequent study by a person or team other than the authors, without the use of author-supplied artifacts” [1]. In principles, new research studies should first reproduce or replicate the work of others, understand and confirm them, and then build improvements from there. Motivations for openness in energy system models include helping public policies and improving public trust as well as improving reproducibility and collaboration within the scientific community [2].

However, the pace of research and “publish or perish” paradigm has made the upkeep of these pillars increasingly difficult, at least in some areas. In the domain of Multi-Energy Systems (MES), research studies are often targeting at very diverse systems, including some unique real systems. In the end, studies with very similar scientific scopes have divergent assumptions and terminologies, as illustrated by Schischke et al. [3], which limits the opportunity for properly validating research results with reproduction and replication. This situation leads to (i) disappointment when trying to apply research results, especially when many aspects have been neglected in order to conclude studies fast enough and with limited budget (ii) many studies being done over and over in slightly different contexts, effectively increasing publication throughput, but limiting the actual progression of knowledge and causing parallel efforts (iii) sometimes similar studies producing different results due to slightly different but undocumented assumptions (on input data, models …).

While full reproducibility and replication may not be possible for studies on MES due to their specificities, the research community should aim at least for studies to be comparable, e.g., by applying approaches and methods to reference case studies or even benchmarks. This paper takes a first step by investigating an approach for structuring existing case studies descriptions in order to make them more comparable.

The paper is structured as follows. Section 1.1 reviews some aspects of the state of the art regarding comparability of studies in MES. Section 2 introduces the method followed by the authors for conducting the study. Section 3 details the obtained results, focusing in particular on the outcomes of the case study review. Section 4 provides discussion, and section 5 concludes this paper.

## 1.1. State of the Art

Before introducing the contribution of this paper, this section investigates the current status of comparability of studies in MES.

### 1.1.1. Benchmarks for Multi-Energy Systems

Benchmarks provide the strongest level of comparability: different methods are applied on the same cases, delivering KPIs that can be compared in order to better understand the difference between the approaches. Established benchmarks in multi-energy systems are scarce, although several efforts recently emerged in this direction [4–6]. On the other hand, established benchmarks exist for specific subdomains, and should be considered as a basis. The most numerous and notable benchmarks were probably developed for the electric-grids. The very first open-access set of distribution test feeders were published in 1991, though many well-established test cases of power systems appeared back in 1960s [7]. Since this time, international working groups IEEE-PES [8,9], CIGRE [10] , system operators as well as many other initiatives have been expanding electric-power benchmarks for: power systems analysis [11], stability analysis [12], distributed energy resources [13], congestion management [14], power quality [15], renewables integration [16]. Benchmarks in gas [17–19] and heating networks [20,21] were also following this trend though seems having much less data and tools available in open access.

In order to better understand the situation, it should also be considered that studies on MES cover a wide range of objectives as well as spatial and temporal scales [22,23]. However, the modeling, simulation and optimization approaches employed in this wide range of studies overlap significantly, and should still be comparable to some extent. Understanding the objectives and hypotheses of all cases included in a benchmark is especially required to understand the impact of simplifying hypothesis when moving to larger time and space scales.

### 1.1.2. Description frameworks for case studies in Multi-Energy Systems

As pointed out by [24], establishing relevant benchmarks in energy modeling requires significant effort. For this reason, a first focus can be to reference case studies which could serve as a starting point. A well-documented case study provides a practical starting point for comparison, by letting other studies adapt the reference to its context and define its own Key Performance Indicators (KPIs).

Describing and sharing a case study so that others can understand it in details or even try to reproduce it is still not straightforward. Previous works have been investigating the way case studies should be documented and described.

Open energy modeling practices emphasize the importance of documenting the entire process—from raw data to interpretation [25] —alongside contextual information such as objectives and assumptions. Based on these recommendations, Hülk et al. [26] defined energy scenarios factsheets describing scenarios, models and tools in the Open Energy Platform. Scenario factsheets are further grouped together as scenario bundles, in order to provide “all relevant information to understand a scenario's context and to ease a potential re-use of quantitative data for your own purposes.” [27].

Another approach comes from the IEC 62559 standard [28], which sets normative requirements for case study documentation in the context of normalization activities. Although strictly adhering to these requirements is time-consuming and mostly inaccessible outside standardization activities, projects such as Smart4RES adapted a subset of IEC 62559 for their own use [29]. However, this was limited to a superficial description of the case study, aimed at pointing out the links between several activities rather than facilitating the reuse of case studies.

Learning from these previous experiences, the ORUCE methodology introduced a possibility for sharing use cases more easily and efficiently. While retaining some of the formal aspects of OEP, ORUCE proposes to use a single template for describing a case study, with sections that globally match the various fields required in the OEP formulars [30]. Furthermore, ORUCE hints at making the description available directly in Jupyter notebooks, so that more detailed aspects of the model may even be directly executable and visualized.

A similar but more detailed effort is the one introduced with the PreCISE (Preparing Concise Information for Simulation Experiments) approach. While ORUCE primarily focuses on optimization studies, PreCISE targets modeling, simulation, and optimization in a more agnostic way. It was designed in the context of the Smiles project [31] and later used in other projects such as EriGrid2 [32,33]. The PreCISE framework provides a structured, tool-agnostic methodology for documenting simulation and optimization studies in multi-energy systems. Its central objective is to harmonize the exchange of information among research teams working with heterogeneous modelling paradigms, simulation toolchains, and optimization approaches. PreCISE achieves this through standardized templates that describe all elements required to implement a simulation experiment, independently of any specific software environment.

## 1.2. Contribution of this paper

While each of the mentioned approaches provide relevant aspects towards making MES case studies more comparable, none of them has been widely adopted at this stage. This paper aims to address this gap by evaluating more precisely how such a description framework can be used on a set of diverse case studies. The contributions of this paper are as follows:

- Making a first step towards reference case studies and benchmarks for researchers working on MES
- Pushing forward a description framework suitable for MES, based on existing literature
- Evaluating the proposed description framework on a set of case studies investigated by the co-authors from various institutes
- Drawing some initial conclusions on the potential for a wider adoption of a standardized description framework
- Making recommendations for evolutions of existing description frameworks in order to fulfill the needs for research in Multi-Energy Systems.

# 2. Materials and Methods

## 2.1. Presentation of the adopted process for conducting the study

The study was conducted through a structured, five-step process:

1. **Framework Analysis**: Some of the authors analysed existing description frameworks, as presented in the state-of-the-art (section 2.2).
2. **Framework Selection**: Based on this analysis, a subset of the PreCISE framework was selected, especially focusing on the system configuration, test case/test specification, and input data.
3. **Case Study Description**: Each co-author described their own MES case study using the selected framework, ensuring a diverse representation of systems, methodologies, and research objectives. The descriptions were shared on Github.
4. **Cross-Review**: Co-authors reviewed each other's case study descriptions, focusing on completeness, clarity, and openness of data and models.
5. **Feedback Integration**: Feedback was gathered to identify strengths and limitations of the framework, with a particular emphasis on opportunities for simplification, standardization of terms, and broader applicability.

## 2.2. Initial analysis of description frameworks

Several criteria should be considered for choosing a description framework for this study:

- **Flexibility**: it should accommodate the description of case studies regarding both simulation and optimization models, or even combinations of both. It should suitable for describing a given case study for different phases of a MES lifecycle (from early planning to actual system deployment).
- **Precision**: while being flexible, it should aim at minimizing ambiguity, especially by providing a well-defined structure and vocabulary. The description framework should facilitate the comparison of similarities and differences between case studies used by different authors.
- **Tool agnostic**: it should not be tied to a specific tool or model. It should abstract the underlying models, for the purpose of communicating details about the case study, not performing computation or formal analysis.
- **Ease of adoption**: it should be designed easily enough to facilitate adoption, especially by providing multiple levels of descriptions which can be enhanced gradually.
- **Scale**: it should support various system sizes (e.g., building, industrial facility, district, city, regional) and enable a detailed description of the energy networks when relevant to the case study.
- **Variations**: it must effectively handle parameter variations, scenarios, as well as alternative system configurations when applicable.

Table 1 provides an initial qualitative analysis of some description frameworks, considering these criteria. Although this analysis would need to be deepened, it led to the selection of the PreCISE description framework, which particularly appeared to strike a good balance between, flexibility and precisions. It should nevertheless be noted that the criteria related to handling parameters variations or scenarios did not seem natural in any of the considered framework, except OEP.

***Table 1.*** *Initial qualitative analysis of some description frameworks*

| Month | Flexibility | Precision | Tool-agnostic | Ease of use | Scale | Variations |
|---|---|---|---|---|---|---|
| OEP | -- | +++ | + | - | + | + |
| ORUCE | ++ | + | + | + | - | ? |
| IEC 62559 | - | +++ | ++ | - | + | ? |
| Smart4RES UC | - | + | ++ | + | + | ? |
| PreCISE | ++ | ++ | ++ | + | + | ? |

## 2.3. Overview of the PreCISE description framework

In the PreCISE framework, the standardized description of case studies relies on three tightly connected elements:

- the system configuration,
- the test case, and
- the test specification.

The system configuration represents the static description of a real or conceptual energy system, including all components, their attributes, interconnections, and contextual characteristics. It provides the technological and structural reference that underlies all subsequent assessments but deliberately excludes control behaviour or use-case-specific dynamics. By formalizing component hierarchies, physical properties and network topologies in a technology-neutral way, it allows different modelling teams to reconstruct functionally equivalent systems even when using incompatible modelling paradigms [31].

A test case formulates the purpose of an assessment. It specifies the functional objective, the object under investigation and the performance criteria but does not prescribe how the system is to be simulated. The test case therefore defines the analytical intent (e.g., characterization, verification or optimization) while maintaining independence from tools, models or assumptions about simulation or optimization setup. This abstraction ensures that multiple research groups can pursue the same analytical question without being constrained by methodological differences.

The test specification is the operational element that maps the conceptual test case onto an executable test system. It defines the concrete simulation or optimization setup by specifying boundary conditions, simplifications, test parameters, input data, expected outputs, temporal resolution, initial conditions, stopping criteria and uncertainty considerations. In contrast to the test case, the test specification describes how a particular experiment is to be carried out, but it still avoids tool-specific implementation details. This separation allows different modelling teams to implement the same test specification in different toolchains while ensuring that their results remain meaningfully comparable.

## 2.4. Checklist for reviewing case studies descriptions

To ensure a consistent and comparable assessment of the submitted case studies, a structured review checklist was developed. Its purpose is to guide reviewers through the evaluation process and to verify whether each case study provides the information necessary for understanding and reproducing the work, regardless of the modelling framework applied. The checklist is organized into five sections: A) Metadata & Context, B) System Configuration, C) Temporal & Spatial Resolution, D) Data, and E) Assumptions & Simplifications. An additional open-comment field allows reviewers to provide any further relevant observations.

Each section contains targeted questions designed to assess the completeness and clarity of the case study description. Reviewers assign a score from 0 to 3, where 0 indicates missing information and 3 indicates that the information is clear, complete, and ready for replication. Reviewers may also select N/A if a question is not relevant to the case study, or F if the description framework does not allow the authors to provide the requested information. Finally, reviewers are asked to justify their ratings with short explanatory notes. The full review template is available on the GitHub repository.

# 3. Results

## 3.1. Overview of contributed case study descriptions

An overview of the contributed case study descriptions is available in Table 3 (next page). Full details about case studies can be found at https://github.com/mathieu-vallee/ecos-case-studies.

## 3.2. Review of case study descriptions

The review process was explicitly designed as a framework validation exercise, not as a benchmarking or comparison of the case studies themselves. Accordingly, the assigned ratings reflect how well the adapted description template enables authors to document essential characteristics of their multi-energy systems, and how effectively reviewers can reconstruct and understand the case studies based solely on the structured descriptions. The scores should therefore not be interpreted as an assessment of the intrinsic quality, sophistication, or relevance of the individual case studies.

In total, 24 reviews were conducted for ten case study descriptions, resulting in two to three independent reviews per case study. All reviews were performed using the structured checklist introduced in Section 2.3. The completed review forms are publicly available via the project's GitHub repository. In the following, the review results are summarized and discussed with a focus on the strengths and limitations of the description framework.

Table 2 summarizes the distribution of reviewer ratings across the five checklist sections. Ratings of 2 (“mostly complete and clear”) and 3 (“clear, complete, and ready for replication”) dominate across all sections, indicating that the framework generally supports structured and interpretable case study documentation. At the same time, noticeable variation between sections and reviewers provides valuable insights into areas where the framework offers insufficient guidance or where expectations are not yet fully standardized.

***Table** 2: Percentage distribution of reviewer ratings by checklist section.*

| Section | 0 | 1 | 2 | 3 | N/A | F |
|---|---|---|---|---|---|---|
| A) Metadata & Context | 2% | 11% | 21% | 64% | 0% | 0% |
| B) System Configuration | 1% | 15% | 26% | 56% | 0% | 0% |
| C) Temporal & Spatial Resolution | 7% | 5% | 11% | 62% | 10% | 2% |
| D) Data | 10% | 22% | 26% | 33% | 4% | 4% |
| E) Assumptions & Simplifications | 15% | 2% | 42% | 33% | 0% | 8% |

Some degree of inter-reviewer variability was expected, as reviewers naturally interpret questions and completeness thresholds differently. Importantly, such divergence is informative: it often highlights ambiguous terminology, missing structure, or conceptual gaps in the framework rather than deficiencies in individual descriptions.

**Section A) Metadata & Context**

This section received consistently high ratings, suggesting that the framework successfully guides authors in describing the general purpose and context of their case studies. Lower ratings primarily stem from missing or unclear metadata, such as licensing information, authorship, or contact details, as well as from insufficient methodological transparency. While the overall motivation and scope are usually understandable, information required for reuse or reproduction is sometimes incomplete or distributed across external sources.

**Section B) System Configuration**

System configuration is generally well described, particularly in textual form. However, reviewers noted inconsistencies between textual descriptions, system breakdowns, and graphical representations. Unclear or non-standardized schematics reduce interpretability, even when the underlying system is conceptually sound. These findings suggest that the framework could benefit from stronger guidance on visual system representation, for example through recommended graphical conventions or a minimal, standardized graph schema (nodes, edges, and attributes).

**Section C) Temporal & Spatial Resolution**

Most descriptions clearly state time horizons and resolution, resulting in a high share of ratings of 3. Nevertheless, reviewers often struggled to fully reconstruct temporal abstraction methods. Concepts such as representative days or weeks are commonly mentioned, but their derivation, aggregation logic, and implications for system boundaries are frequently underspecified. This section also exhibits the highest share of “N/A” and “F” ratings. While “N/A” often reflects the heterogeneity of modelling approaches, the “F” ratings point to a structural limitation: the framework currently lacks a dedicated and standardized place to describe clustering and temporal reduction methodologies.

*Table 3. Overview of the case study descriptions provided by the co-authors to evaluate the description framework.*

| Case ID | Name | Institution | Sector/ system type | Test objective | Model | System description | Key publications |
|---|---|---|---|---|---|---|---|
| MENB | Multi-energy network benchmark | AIT | Electrical network, thermal network | Characterization | Dynamic | Typical Central European city | [32] |
| GRCAM | Grenoble Cambridge district | CEA | Residential, heating networks | Optimization/ simulation / characterization | Dynamic + MILP | Residential district in the city of Grenoble, France | [34–36] |
| DH-EG-Storage | Integrated electrical and thermal network with storage | IMT | Heating network, electrical grid, energy hubs | Optimization | Dynamic/static | Coupled district heating network and electrical grid integrating coupling technologies (CHP, HP) renewable generation (Wind and Solar) and thermal storage | [37] |
| SC-SunSTONE-CHB-2019 | Châteaubriant district heating network | INSA | Residential, heating networks | Characterization | Dynamic | District heating network in Châteaubriant, France | [38] |
| POLITO-NITDEH | Northern Italy district energy hub | Polito | Residential, heating networks | Optimization | GA + dynamic DH model + LP | District energy hub in Northern Italy | [39] |
| MMEH | Munich multi-energy hub | RWTH | Mixed-use, heating network, cooling networks, electrical network | Optimization | MILP | District energy system in Munich, Germany | [40,41] |
| SPICS | Steel processing industry case study | TU Wien | Industry, industrial site | Optimization | MILP | Industrial steel processing site in Judenburg, Austria | [42] |
| BF-MFH-C-Q45Strom | Begleitforschung Multi-family homes | UMSICHT | Residential, heating networks | Optimization | MILP | Synthetic Multi-family homes in Potsdam, Germany | [43] |
| DOMES | University district in Padova | UNIPD | Tertiary, district level, electrical and thermal network | Optimization | MILP | University district in Padova, Italy | [44,45] |
| UNIPR-Campus | University of Parma Campus | UNIPR | Service sector heating and cooling network | Simulation/ Characterization | Dynamic/static | Campus of University of Parma, Italy | [46] |

**Section D) Data**

The data section shows the widest spread of ratings and the highest shares of low scores. This is partly explained by the framework adaptation process used itself, as data-related aspects were added at a later stage and received less emphasis. Across reviews, the dominant issues are limited data accessibility, missing quantitative detail, and insufficient traceability. Units, conventions, data sources, preprocessing steps, and links to repositories are unclear or missing. These results indicate that the framework (as it was used for this paper) does not yet sufficiently structure data documentation, despite data being essential for reproducibility. At the same time, this section is inherently challenging, as relevant information is often scattered across code, input files, and external datasets.

**Section E) Assumptions & Simplifications**

This section reveals systematic challenges. Reviewers frequently noted insufficient transparency regarding modelling assumptions, particularly with respect to control strategies, optimization formulations, investment logic, and solver settings. The relatively high share of “F” ratings indicates that the framework provides limited support for capturing such information. Many assumptions are implicitly embedded in modelling tools or code bases and are therefore difficult to describe using the current template structure.

Reviewers were provided only with the adapted description template and did not receive additional calibration or training. As a result, the review outcomes reflect genuine first impressions and natural ambiguities, which makes them particularly valuable for identifying usability issues and potential for improvement.

Overall, the review confirms that the adapted framework is largely effective for documenting multi-energy system case studies in a structured and comparable way. Reviewers were generally able to reconstruct key aspects of system configuration, temporal structure, and modelling scope. However, several categories reveal systematic limitations that point to opportunities for refinement. In particular, future iterations should provide stronger guidance for (i) temporal abstraction and clustering methods, (ii) data provenance and preprocessing, and (iii) modelling assumptions that are implicitly embedded in optimization or simulation environments.

Finally, reviewers and authors alike noted that completing the full framework requires a considerable time investment. While this effort significantly enhances transparency and reusability for the scientific community, it competes with tight project and publication deadlines. Providing incentives, tooling support, or partial automation for framework completion could therefore be crucial for broader adoption.

# 4. Discussion

The main goal of such a framework is to describe multi-energy system case studies in a sufficiently clear and structured manner to support understanding, reuse, and meaningful comparison across studies. As discussed in Section 2.2, achieving this goal requires the framework to satisfy several complementary classes of requirements: scientific requirements related to clarity, precision, and comparability; methodological requirements related to flexibility, scale, and tool agnosticism; and practical requirements related to ease of adoption, communication, and long-term reuse.

A structured description framework should be regarded not only as a means to improve comparability, but as a prerequisite for moving MES studies toward the level of artifact maturity expected in other research communities, where case studies and associated resources are sufficiently documented, reusable, and open to independent validation.

From this perspective, the results suggest that the adopted framework provides a credible basis for structuring MES case study descriptions, while also revealing several limitations that are directly relevant to its further development. Overall, the review outcomes indicate that the adapted framework improves the interpretability and first-order comparability of MES case studies by making objectives, system configuration, and general modelling choices more explicit. However, its current use as a basis for reuse, rigorous comparison, or reconstruction remains limited by gaps in the documentation of data, assumptions, and methodological abstractions. This is shown by more dispersed ratings observed for these aspects.

Regarding scientific requirements, the predominance of ratings of 2 and 3 across most checklist sections indicates that reviewers were generally able to reconstruct the purpose, scope, and main characteristics of the case studies from the structured descriptions alone. This is an important result, because it suggests that the framework succeeds in reducing ambiguity and supporting a first level of understanding and comparability of MES studies. However, the results also show that comparability remains partial when essential elements such as data provenance, pre-processing steps, or modelling assumptions are insufficiently documented. Another recurring limitation was that the framework does not always clearly distinguish between information that is missing, intentionally omitted, or not relevant to a particular case study. This ambiguity reduces the reviewers' ability to assess completeness consistently and complicates cross-case comparison.

Concerning methodological requirements, the diversity of the contributed case studies demonstrates that the framework is flexible enough to accommodate different system types, modelling paradigms, and spatial scales. This supports the choice of a tool-agnostic and relatively general structure. At the same time, the review suggests that its underlying hierarchical and object-oriented logic is more natural for some modelling paradigms than for others. In particular, highly aggregated or non-object-oriented approaches may be more difficult to document within the current structure. At the same time, some methodological aspects appear insufficiently formalized, especially the description of temporal abstraction methods and assumptions embedded in optimization or simulation environments. These are precisely the areas where methodological heterogeneity in MES can strongly affect results, and where the framework would need to provide more explicit guidance.

With respect to practical requirements, the study also highlights a central tension. On the one hand, the framework enables a richer description of case studies making them more reusable, which is a clear benefit for the community. On the other hand, both authors and reviewers noted that completing the description remains time-consuming, especially under typical publication constraints. Considering that ease of adoption is not a secondary usability issue, but a key condition for whether such a framework can realistically contribute to broader standardization and reproducibility, reducing effort and improving guidance is central. Clearer templates, explicit checklists of required information, and more standardized graphical conventions could reduce interpretation effort for both authors and reviewers and make the framework easier to use consistently.

Table 4 (next page) summarizes the key findings. Taken together, these findings suggest that the adapted framework already fulfils an important part of its intended role: it improves the structured description and sharing of MES case studies and provides a more explicit basis for their understanding and comparison. In that sense. While the framework does not by itself guarantee reproducibility, replication, or benchmarking, it contributes to the conditions under which these objectives become more realistic. Hence, it paves the way for the development of shared reference cases and, in the longer term, more robust benchmarking practices within the MES research community.

# 5. Conclusion

Research on Multi-Energy Systems (MES) often relies on case studies with divergent hypotheses and terminologies, limiting comparability and slowing progress. In order to improve comparability, this paper investigates the use of a standardized description framework for MES case studies. This can be considered as a first step towards defining reference case studies and benchmarks for MES.

The study adopted a structured, five-step collaborative process: (i) analysis of existing description frameworks, (ii) selection and adaptation of the PreCISE framework, (iii) description of MES case studies by co-authors from different institutes, (iv) cross-reviews of these descriptions, and (v) integration of feedback to refine the framework. This process resulted in 10 case study descriptions and 24 independent reviews, providing robust insights into the framework's strengths and areas for improvement.

The study underscored the PreCISE framework's effectiveness in addressing key scientific requirements, such as improving clarity, precision, and comparability. This was particularly the case for metadata, general system configuration and topology, and the description of basic temporal and spatial characteristics. However, it also revealed some limitations in documenting data provenance, pre-processing steps, and embedded modelling assumptions, which were often not completely defined by case study providers.

The diversity of case studies also confirmed the suitability regarding methodological requirements related to flexibility, scale, and tool agnosticism. Nevertheless, some methodological aspects appear insufficiently formalized, especially the description of temporal abstraction methods and assumptions embedded in optimization or simulation environments. These are precisely the areas where methodological heterogeneity in MES can strongly affect results, and where the framework would need to provide more explicit guidance.

*__Table__ 4:* *Summary of the PreCISE description framework's satisfaction of requirements*

| **Requirement type** | **Expected property** | **Status** | **Main comments** |
|---|---|---|---|
| Scientific | Clarity | Good | Reviewers were able to reconstruct the purpose, scope and main characteristics of case studies. |
| | Precision | Variable | Essential elements such as data provenance, pre-processing steps, or modelling assumptions are insufficiently documented. |
| | Comparability | Variable | The framework does not always clearly distinguish between information that is missing, intentionally omitted, or not relevant to a particular case study |
| Methodological | Flexibility | Good | Flexible enough to accommodate different system types, modelling paradigms, and temporal and spatial scales. |
| | Scale | Good | Case studies at different temporal and spatial scales can be described |
| | Tool-agnosticism | Good | Some methodological aspects appear insufficiently formalized, especially the description of temporal abstraction methods and assumptions embedded in optimization or simulation environments |
| Practical | Ease of adoption | To be improved | Completing the description remains time-consuming, especially under typical publication constraints. Reducing effort and improving guidance is central. |
| | Communication | To be improved | Clearer templates, explicit checklists of required information, and more standardized graphical conventions could reduce interpretation effort |
| | Long-term reuse | Good | Enables a richer description of case studies making them more reusable |

Finally, the study indicates that more work is required regarding practical requirements. While the framework did facilitate the adoption and participation of all co-authors, the study also highlighted the tension between the time investment required for thorough documentation and the practical constraints of research timelines. In this domain, improving tooling and especially more structured adoption of relevant generative AI tools could provide significant help for researchers.

While the framework does not guarantee full reproducibility or benchmarking, it significantly advances the structured sharing and understanding of MES case studies, creating a foundation for more mature, reusable, and comparable research artifacts. Future efforts should focus on (i) improving guidance for using the description framework, for instance prioritizing required information depending on context and case study maturity (ii) reducing ambiguity on some aspects, especially by providing pre-defined terminologies & definitions (iii) provide tooling for helping to fill the description.

Future efforts should focus on refining guidance for temporal abstraction, data documentation, and assumption transparency, as well as exploring incentives or automation to facilitate broader adoption. By fostering clearer communication and reducing ambiguity, this work represents a meaningful step toward standardized, collaborative progress in MES research.

# Acknowledgments

M. Vallee would like to particularly thank Prof. Bruno Lacarrière for initial discussions leading to the proposal of this paper, as well as Daniel Muschick from BEST research for his participation in the initial stage. The work of M. Vallee, I. Daminov, M.-T. Mabrouk and S. Hodencq has been founded by the National French Research Agency (ANR) in the PEPR TASE HyMES project, under grant number 22-PETA-0002.